\documentclass[12pt,letterpaper]{article}
\usepackage{amsmath,amssymb}
\usepackage{graphicx,color}
\usepackage[linkcolor=blue,colorlinks=true]{hyperref}
\usepackage{cite,epsf}
\usepackage[bf]{caption}
\usepackage{subfig}
\usepackage{mathrsfs}

\def\a{\alpha}
\def\b{\beta}
\def\g{\gamma}

\def\e{\epsilon}

\def\k{\kappa}
\def\l{\lambda}
\def\L{\Lambda}
\def\m{\mu}
\def\n{\nu}

\def\r{\rho}

\def\s{\sigma}

\def\t{\tau}

\def\o{\omega}
\def\O{\Omega}

\def\be{\begin{equation}}
\def\ee{\end{equation}}
\def\bea{\begin{eqnarray}}
\def\eea{\end{eqnarray}}
\def\ba{\begin{aligned}}
\def\ea{\end{aligned}}

\def\12{\frac{1}{2}}

\def\eNm1{\overset{\scriptscriptstyle{(N-1)}}{e}}

\usepackage{xargs}
\usepackage[colorinlistoftodos,prependcaption,textsize=tiny]{todonotes}
\newcommandx{\Stefan}[1]{\todo[backgroundcolor=red!25,bordercolor=blue,noline]{St:#1}}
\newcommandx{\Sadik}[1]{\todo[backgroundcolor=blue!25,bordercolor=red,noline]{Sa:#1}}

\begin{document}

\title{
\hfill\\
\hfill\\
A Review of Third Way Consistent Theories
\\[0.5cm]
}

\author{
Nihat Sadik Deger
}

\date{}

\maketitle
\vspace{-1.5cm}

\begin{center}
\vspace{0.5cm}\textit{\small
Department of Mathematics, Bogazici University,\\
Bebek, 34342, Istanbul, Turkey
}

\end{center}
\vspace{5pt}

\abstract{We will give an overview of ``third way consistent" theories. Field
equations of such models do not come from the variation of a local action
without auxiliary fields, yet their covariant divergences still vanish on-shell.
First examples were discovered in three dimensions which were pure massive
gravity and Yang-Mills theories. However, recently interacting $p$-form theories
with this property in arbitrary dimensions were also constructed. 
After explaining construction of these theories
and some of their general features, we will discuss some open problems and future directions.}

\noindent

\thispagestyle{empty}

\newpage

\section{Introduction}

The action principle has undeniably been central in the development of modern physics. Equations that  
describe 4 fundamental forces of our universe, namely those of General Relativity
and the Standard Model, can be derived using this principle. Having an action helps in understanding
interactions, symmetries and quantizing the theory. However, there are some theories for which
no simple action exists. A well-known example is the 10-dimensional Type IIB supergravity which contains a 4-form field with a
self-dual field strength. As a result, its equations of motion can not be derived from a 
Lorentz invariant action without extra fields \cite{Marcus:1982yu}. Similarly, there is no action
for supergravities in which the global scaling symmetry is gauged \cite{LeDiffon:2008sh}. These models
are defined via their
field equations whose consistency works as usual. In principle, there might also be 
some physical phenomena or theoretical questions whose explanations require inclusion of interaction terms in the field equations 
that do not come from an action without auxiliary fields and Bianchi identities are satisfied in an unusual way. Such a situation occurred in 3-dimensions
where search for a pure higher curvature gravity theory which has both bulk and boundary unitarity led to the discovery
of the Minimal Massive Gravity (MMG) model \cite{Bergshoeff:2014pca}. The novelty of its modified Einstein's equation
is that, its covariant divergence vanishes only after using the Einstein's equation again. This rather surprising way of establishing consistency is called the ``third way". Other such
gravity \cite{Ozkan:2018cxj, Afshar:2019npk} and gauge theory \cite{Arvanitakis:2015oga} examples in 3-dimensions were found later. Recently, such $p$-form theories in arbitrary dimensions are built in \cite{Broccoli:2021pvv}. Moreover, in \cite{Deger:2021fvv} $N=1$ supersymmetric version of \cite{Arvanitakis:2015oga} is obtained. 

In this paper we will focus on the construction of this new type of theories
and leave other important issues, such as their exact solutions, couplings etc. aside. One common feature they
all share in their construction is the vital role played by a shift in the connection of the model.
As will be seen, in principle
it is possible to obtain more examples in three dimensions using the methods that we will describe. However, finding new models in higher dimensions is a challenge. We will start by defining the term ``third way" in the
next section which is also reviewed in \cite{Bergshoeff:2015zga}. Constructions of known gravity and gauge models with this property are explained in sections \ref{gravity} and \ref{gauge} respectively. We conclude in section \ref{future}
by indicating some future directions.

\section{What is the ``Third Way"?}

To explain what is meant by the "third way" let us consider the vacuum Einstein's equation:
\be
G_{\m\n} \equiv R_{\m\n} - \frac{1}{2} Rg_{\m\n}=0 \, ,
\label{Einstein}
\ee
which can be derived from the Einstein-Hilbert action. Any addition
to this equation should be compatible with the fact that the divergence of the Einstein tensor $G_{\m\n}$
vanishes identically. There are two familiar and one uncommon sorts of modifications:

\

{\it
i) Adding divergence free gravitational terms:} 
Historically, it was Einstein himself who first considered such a modification by introducing a cosmological
constant $\Lambda$ which also follows from an action. In 3-dimensions a famous generalization is the (cosmological)
topologically massive gravity (TMG) \cite{Deser:1981wh} whose field equation is:
\be
G_{\m\n} = -\Lambda g_{\m\n} -\frac{1}{\m} C_{\m\n} \, ,
\label{tmg}
\ee
where $\m$ is a mass parameter. The Cotton tensor $C_{\m\n}$ is defined as
\be
C_{\m\n} \equiv \epsilon_{\m\r\s}\nabla^\r S^\s_{\,\,\,\,  \n} \, ,
\label{cotton}
\ee
where $S_{\m\n}\equiv G_{\m\n} + \frac{1}{4}Rg_{\m\n}$ is the Schouten tensor. It is straightforward to show that
the Cotton tensor is covariantly conserved. Indeed this is not a surprise, since any additional gravitational
term to \eqref{Einstein} that comes from a diffeomorphism invariant action will automatically be divergent free and
the Cotton tensor can be obtained from the variation of the gravitational Chern-Simons action.

\

{\it ii) Adding other fields:}
Let us now consider the following modification of \eqref{Einstein}:
\be
G_{\m\n} = \k (F_{\m\r} F^{\r}_{\,\,\,\, \n} - \frac{1}{4}g_{\m\n} F_{\r\s}F^{\r\s}) \, ,
\label{max1}
\ee
where $F_{\m\n}\equiv \partial_\m A_\n -\partial_\n A_\m$ is the field strength of an abelian vector field $A_\m$
and $\k$ is a coupling constant. If we now take the divergence of this equation we see that the r.h.s vanishes only when
\be
\nabla_\a F^{\a\b}=0 \, .
\label{maxwell}
\ee
Actually \eqref{max1} and \eqref{maxwell} are nothing but equations of motion of the Einstein-Maxwell action.
In general, any covariant coupling of a matter action to the Einstein-Hilbert action will change the gravitational 
field equation as $G_{\m\n} = \k T_{\m\n}$ where $T_{\m\n}$ is the matter energy-momentum tensor which
is conserved when matter field equations are satisfied by the Noether's theorem.

\

{\it iii) Adding gravitational terms \`a la third way:} Now consider the following generalization of the TMG field equation \eqref{tmg}:
\be
G_{\m\n} = -\Lambda g_{\m\n} -\frac{1}{\m} C_{\m\n} -\frac{\g}{\m^2} J_{\m\n} \, ,
\label{mmg}
\ee
where $\g$ is a dimensionless constant and the new piece is given by \cite{Bergshoeff:2014pca}:
\be
J_{\m\n} \equiv \frac{1}{2} \epsilon_{\m\r\s}\epsilon_{\n\a\b} S^{\r\a} S^{\s\b} \, .
\label{Jtensor}
\ee
Since the tensor $J$ is built out of metric and curvature tensors only, for this to be a valid modification,
one would expect its divergence to vanish identically as in the class i) that we considered above. Instead, one derives:
\be
\nabla_\m J^{\m\n}= \e^{\n\a\b} S_\a^{\,\,\,\, \l} C_{\l \b} \, . 
\label{divj}
\ee
This shows that the $J$-term can not come from variation of an action of the metric field alone and hence one may conclude
that equation \eqref{mmg} is inconsistent. However, note in \eqref{divj} that the divergence of $J$ is proportional to
the Cotton tensor which appears in the main field equation \eqref{mmg}. Now replacing the Cotton tensor in \eqref{divj}
using \eqref{mmg} one gets:
\be
\nabla_\m J^{\m\n}= -\m \e^{\n\a\b} S_\a^{\,\,\,\, \l} \left( S_{\l\b} - \frac{1}{4}R g_{\l\b} + \Lambda g_{\l\b} + \frac{\g}{\m^2} J_{\l\b} \right) \, , 
\ee
which is zero since each term vanishes. Note in particular that, $SJ \sim S^3 =0$ due to the way indices are arranged. Hence, we conclude that equation \eqref{mmg} makes sense on-shell and the model it describes is called the Minimal Massive
Gravity (MMG) \cite{Bergshoeff:2014pca}. Actually, here we have a 'mixture' of the two cases that we discussed above;
namely we have a purely gravitational modification (as in case i)) but we have to use the gravity equation itself again (as in case ii)) for consistency. This is what is meant by the "third way" and as we will see it is not something particular to
gravity; such gauge theories also exist.

\section{Gravity Examples} \label{gravity}
Coming up with a third way consistent gravity equation such as \eqref{mmg} is highly non-trivial. That is probably why this class of theories were not discovered until recently. Their origin in 3-dimensions becomes more transparent using
the Chern-Simons-like description of gravity developed in \cite{Hohm:2012vh, Bergshoeff:2014bia, Afshar:2014ffa, Merbis:2014vja}
which we summarize below.\footnote{Construction of  such models directly in the metric formulation is described in \cite{Alkac:2018eck}.}

In this formulation dynamical fields ($X_K$'s with $K=0,1,2,, ...$) are a set of Lorentz-vector valued 1-forms $X_K^a= X_{K\m}^adx^\m$  where the generators of the SO(1,2) Lorentz algebra satisfy $[J_a, J_b]= \epsilon_{abc}J^c$. The index $K$ indicates the weight of the field and we split those with even and odd weight as $f_{I}$ (with weight $2I$) and $h_{I}$ (with weight $(2I+1)$) respectively. Moreover, we define $f_0\equiv e$ (with weight zero) as the dreibein and $h_0 \equiv \o$ (with weight 1) as the (dual) spin-connection ($2\o^a=  \epsilon^{abc} \o_{ab}$). So, the set of fields that will be used in this construction ordered with increasing weight are: $(e, \o, f_1, h_1, f_2, h_2, .....)$. To build an action we have the following rules:

\

$\bullet$ Lagrangian should be a 3-form.

$\bullet$ Derivatives should be covariant using the spin connection: $ D \equiv d + \frac{1}{2}[\o,]$. Covariant derivative
has weight 1.

$\bullet$ The spin connection $\o$ should appear only in covariant derivatives and in the Chern-Simons combination:
$<\o \wedge d\o + \frac{2}{3}\o \wedge \o \wedge \o>$, where the bracket notation stands for the matrix trace.

$\bullet$ The vanishing of the torsion ($De=0$) should be among the equations of motion. We also assume that the dreibein is invertible. These two together imply that $\o$ is solvable in terms of $e$.

$\bullet$ Equations of motion should be integrable in the sense that each field should 
be solvable in terms of the lower weight fields or their covariant derivatives. In other words, it should be possible to order the equations with increasing weight starting with $De=0$ (which is weight 1) such that the weight $p$ equation has the form:
$DX_{p-1} + e\wedge X_p =0$ \footnote{This equation may also include terms with weight less than $p$.} which can easily be
solved for $X_p$. This means that the whole equation system can be solved recursively until the last one. Finally,
using solutions for the auxiliary fields in the highest weight equation will give the gravity field equation expressed in terms of the metric.

\

Note that the last item in the list restricts possible actions quite strongly. The allowed Lagrangian terms with the above rules up to weight 3 are as follows:

\

\noindent Weight 0: $ e\wedge e\wedge e$ \\
Weight 1: $e\wedge De$ \\
Weight 2: $e\wedge e\wedge f_1$, $ e\wedge R$ (where $R^a=D\o^a$ is the dualized curvature 2-form.)\\
Weight 3: $<\o \wedge d\o + \frac{2}{3}\o \wedge \o \wedge \o>$, $f_1 \wedge De$ (equivalent to $Df_1 \wedge e$), $e \wedge e \wedge h_1$  
 
\

Actions with only even (odd) weight terms are called parity even (odd). A parity even (odd) action $S_k$ has terms
with weight $(k+2)$ or less and should be built out of fields with weight less than $(k+2)$. With this terminology
the first parity even action $S_0$ includes 3-forms with weights 2 or 0 which are built out of $e$ and
$\o$:
\be
S_0 = -\frac{1}{\kappa^2}\int <e\wedge R - \frac{\L}{3} e\wedge e\wedge e> \, ,
\ee 
where $\k$ is a constant. From this action we get $G_{\m\n} +\L g_{\m\n}=0$ as 
the metric field equation.
Similarly, the first parity odd action which is compatible with our rules is:
\be
S_1= \frac{1}{2\kappa^2\m}\int <\o \wedge d\o + \frac{2}{3}\o \wedge \o \wedge \o + 2f_1 \wedge De> \, ,
\ee
which leads to the metric field equation $C_{\m\n}=0$. As the weight of the actions get bigger the number of possible terms
that can be included grows rapidly and figuring out which ones should be kept gets involved. However,
in \cite{Afshar:2014ffa} a systematic construction of $S_{k+2}$ from $S_k$ is given
and since $S_0$ and $S_1$ are known this gives an infinite number of massive gravity models which are free
of scalar ghosts. Of course, it is also possible to consider parity violating actions
by mixing parity odd and even ones. 
The first such action that leads to an integrable system of equations is
\be
S_{ \text{\tiny TMG}}= S_0 + S_1 \, ,
\label{act_tmg}
\ee
which gives rise to the TMG field equation \eqref{tmg}.

Clearly MMG model \eqref{mmg} can not be obtained from such a construction since in the above all auxiliary fields
are solvable in terms of the dreibein and after solving, we can back substitute them in the auxiliary action
to get a metric action. Hence, to get the MMG equation \eqref{mmg} we need to relax some of our rules. Let us now deform
the TMG action \eqref{act_tmg} as follows \cite{Bergshoeff:2014pca}:
\begin{align} \label{mmgaction}
    S_{\text{\tiny MMG}}=S_{ \text{\tiny TMG}} +\frac{\alpha}{\kappa^2\mu^2}\int \langle e\wedge f_1\wedge f_1\rangle \, , 
\end{align}
where $\alpha$ is a dimensionless constant. Note that with this addition we are not increasing the number of auxiliary fields
since $f_1$ was already in $S_1$ but the weight of this additional action is 4 which was not allowed by our rules.
From \eqref{mmgaction} one obtains
\begin{eqnarray}
\delta f_1\qquad&& D e + \tfrac{\alpha}{\mu}\, [e,f_1]= 0\,,  \nonumber\\
\delta \omega\qquad&& \tfrac{1}{\mu}\left(R +  [e, f_1]  \right)-D e=0\,,\label{MMG1}\\
\delta e\qquad&& \tfrac{1}{\mu}D f_1+\tfrac{\alpha}{2\mu^2} [f_1,f_1]-R +\tfrac{1}{2}\Lambda [e,e]=0 \nonumber  \,.
\end{eqnarray}
Obviously here the torsion is non-zero but if we shift the spin connection as 
\begin{align} \label{shift_grav}
    \omega\to\omega -\frac{\alpha}{\mu} f_1 \, , 
\end{align}
equations \eqref{MMG1} become \cite{Bergshoeff:2014pca}:
\begin{eqnarray}
&& D e = 0\,,  \nonumber\\
&& R +(1+\alpha)^2 [e,f_1 ] +\tfrac{\alpha}{2}\Lambda [e,e]=0\,,\label{MMG2} \\
    &&\tfrac{1+\alpha}{\mu}D f_1-\tfrac{\alpha(1+\alpha)}{2\mu^2}[f_1,f_1]-R+\tfrac{1}{2}\Lambda [e,e]=0\,, \nonumber
\end{eqnarray}
where $D$ and $R$ are now defined with respect to the shifted connection. Note that, there is no torsion any more and
assuming $\a \neq -1$ the second equation
can be solved for $f_1$ as $(f_1)_{\m\n}=-(S_{\m\n} + \a \L g_{\m\n}/2)/(1+\a)^2$.
Substituting this to the last equation above we get the MMG field equation \eqref{mmg}
with $\gamma=-\a/(1+\a)^2$. The extra term $[f_1,f_1]$ is the source of the $J$-tensor \eqref{Jtensor}.

The above shows that we can relax the no torsion condition in this construction. It is enough to have
an integrable system of equations after a linear shift of the connection with auxiliary fields. However, 
after this shift, back substitution of solved fields to the auxiliary action is not allowed any more and hence there is no
metric action for such a gravity model. 

Although this procedure gives a valuable insight about the origin of MMG,
it is far from obvious how to construct other such models. The next example was found almost
4 years after MMG in \cite{Ozkan:2018cxj} by systematically analyzing the most general Chern-Simons like action
for the fields $(e,\o,f_1,h_1)$ with no weight restriction for the action. Requiring integrability after the shift of the connection (which is with $h_1$)
they found such a first order action for these fields which is parity odd. Its field equation is fourth order in
derivatives of the metric. It is clear that this approach 
gets very complicated as the number of auxiliary fields increases. In \cite{Ozkan:2018cxj} 
initially there are 20 possible terms in the action. In \cite{Afshar:2019npk} it was shown that this
model can be obtained with less effort starting from the $S_3$ action constructed in \cite{Afshar:2014ffa} and by truncating 
the highest weight field $f_2$ as $f_2 \rightarrow c_1f_1 + c_2e$ where $c_i$'s are constant. In general
\begin{align}
    S_{2N+1}[f_{N+1}\to (f_N,\cdots,e)]\longrightarrow \tilde{S}_{2N}\, .
\end{align}
After the truncation the dynamical terms in $\tilde{S}_{2N}$ are fixed. Note that $\tilde{S}_{2N}$ is parity odd.
For the integrability, one may also need to add some parity odd non-dynamical terms which might
violate the weight constraint. However, they are easy to guess once all the dynamical terms are known.
In \cite{Afshar:2019npk} this top-down approach was used to construct another third-way consistent model
whose field content is $(e,\o,f_1,h_1,f_2,h_2)$ starting from $S_5$. Its metric field equation contains sixth order 
derivatives.

\section{Gauge Theory Examples} \label{gauge}
We start with 3-dimensions and then continue with higher dimensional examples. 
Let $A^I_\m$ be a Yang-Mills (YM) gauge field that transforms in the adjoint representation of an
an arbitrary gauge group $G$ with structure constants $f^I_{JK}$. Its field strength is defined as
\begin{equation}
  \label{eq:covquantities}
  F^I_{\mu\nu} \equiv 2 \partial_{[\mu} A_{\nu]}^I + f^I_{JK} A^J_\mu A^K_\nu \,.
\end{equation}
Following \cite{Arvanitakis:2015oga} let us consider the following YM field equation in 3-dimensions:
\begin{align}
\label{MYM}
\epsilon_\mu{}^{\nu\rho} D_\nu \tilde{F}^I_\rho + \mu \tilde{F}^I_\mu = - \frac{1}{2m} \epsilon_\mu{}^{\nu\rho} f^I_{JK} \tilde{F}^J_\nu \tilde{F}^K_\rho  \,, 
\end{align}
where $\mu$ and $m$ are mass parameters and $\tilde{F}_\mu^I \equiv \frac12 \epsilon_\mu{}^{\nu\rho} F^I_{\nu \rho}$ is the
dual field strength\footnote{
This equation with some special choices for the mass parameters appeared earlier in \cite{Mukhi:2011jp, Nilsson:2013fya}.}. The gauge covariant divergence of the l.h.s vanishes automatically thanks to the fact that 
$[D_\m, D_\nu] \tilde{F}^I_\rho \sim f^I_{JK} F^J_{\m\n} \tilde{F}^K_\rho$ and due to the YM Bianchi identity 
$D^\m \tilde{F}^I_\mu =0$. This, of course is not a surprise since these terms come from the YM and Chern-Simons
actions respectively and hence they should be conserved. Therefore, for equation \eqref{MYM} to be consistent r.h.s should
also be conserved. However, it is easy to convince oneself that this term can not come from variation of an
action for $A^I_\m$ alone. Indeed, taking its divergence we see that it is not immediately zero but using \eqref{MYM} again 
we find
\begin{align}
D_\m (\epsilon^{\m \nu\rho} f^I_{JK} \tilde{F}^J_\nu \tilde{F}^K_\rho) \sim 
f^I_{JK} (\epsilon^{\m \nu\rho} D_\m \tilde{F}^J_\nu)\tilde{F}^K_\rho
\sim f^I_{JK} f^J_{MN} \epsilon^{\m \nu\rho} \tilde{F}^M_\m \tilde{F}^N_\n \tilde{F}^K_\rho \, ,
\end{align}
which vanishes due to the Jacobi identity $f^I_{J[K}f^J_{MN]}=0$. 
In \cite{Arvanitakis:2015oga} also an auxiliary action was constructed for this model. However, it was not clear how
to find other such examples as was done in the gravity case \cite{Ozkan:2018cxj, Afshar:2019npk} after MMG \cite{Bergshoeff:2014pca}. This issue was resolved in \cite{Broccoli:2021pvv} which we explain now. To do this we switch to differential form notation for convenience. We begin with the Chern-Simons field equation:
\be
F(A)\equiv dA+A^2=0 \, ,
\label{cs}
\ee 
and shift the connection with a Lie algebra valued 1-form $C$
\be
A \rightarrow A+C \, .
\label{shift}
\ee
We declare what we obtain after this to be our field equation. That gives,  
\be\label{F(AB)}
F(A+C)=F(A)+D_A C+C^2=0 \, ,
\ee
where the covariant derivative of $C$ is $D_A C\equiv dC+AC+CA$. Now \eqref{F(AB)} is our new field equation and
for consistency $D_A F(A+C)=0$ must hold. To see this we use the Bianchi identity $D_AF(A)=0$ 
after which we get
\be\label{con1}
D_A F(A+C)= D_A^2 C+D_A C^2=[F(A),C]+[D_A C,C]=-[C^2,C] =0    \, ,
\ee
where we used the field equation \eqref{F(AB)} in the last step. Hence, we see that equation \eqref{F(AB)} is
third way consistent for any $C$. Moreover, if $C$ is conserved, that is $D_A *\!C=0$, then 
we can generalize it as
\be\label{F(AB)t}
F(A)+D_A C+C^2+\tau *\!C=0   \, ,
\ee
where $\t$ is an arbitrary constant. Equation \eqref{F(AB)t} is again third way consistent due to $[*C,C]=0$.
If we choose $C=\tilde{F}/m$ and $\t=(m-\m)$ then this equation becomes \eqref{MYM}. 
But any other choice of
$C$ will be a generalization and hence this construction gives infinitely many third way consistent models. Moreover,
if $C$ has a conserved part $M$ and non-conserved part $N$, that is $C=M+N, D*\!M=0, D*\!N\neq 0$, then we can modify \eqref{F(AB)}
as
\be\label{F(AB)t2}
F(A)+D_A C+C^2+ c_1*\!M + c_2*\!N=0   \, , 
\ee
where $c_1$ and $c_2$ are constants, if
\be
c_2D*\!N=(c_1-c_2)[*\, N , M] \, .
\ee

Until now we were assuming the spacetime dimension to be three, however in the above construction this did not
play any role except when we moved from \eqref{F(AB)} to \eqref{F(AB)t} which requires the Hodge dual of $C$ to be 
a 2-form. Moreover, the particular choice $C=*F$ is only possible in 3-dimensions. 
Giving up on these, it is straightforward to generalize \eqref{F(AB)} to higher dimensions. For that purpose, let
$B^I$'s be Lie algebra valued $(d-2)$-forms in $d \geq 3$ dimensions  whose YM covariant field strengths defined as
$H^I\equiv DB^I\equiv d B^I+f^I_{JK}A^J B^K$. Note that the dual field strength, denoted by $\tilde{H}$, is a 1-form. Hence,
we can choose $C=\k \tilde{H}$ in the shift \eqref{shift} where $\k$ is a constant and \eqref{F(AB)} becomes
\be
F(A)^I+\kappa\, D_A \tilde{H}^I = -\frac{1}{2} \kappa^2 f^I_{JK}\,\tilde{H}^J \wedge \tilde{H}^K \, ,
\label{gauged}
\ee
which is third way consistent by construction. However, compatibility of YM and $(d-2)$-form gauge symmetries require
$F(A)^I=0$ \cite{Broccoli:2021pvv}. Hence, we set $A^I=0$ in \eqref{gauged} and end up with
\be\label{eom1}
d\tilde H^I=-\tfrac{1}{2}{\kappa}  \, \tilde H^J \wedge \tilde H^K\,f^I_{JK} \, , 
\ee
with $H^I=dB^I$.\footnote{In 3-dimensions $B^I$ is a vector field and \eqref{eom1} can be thought as the
ungauged version of \eqref{MYM}.}
If we hit this equation with the exterior derivative $d$, the l.h.s is zero since it is exact and the r.h.s also vanishes on shell applying the Jacobi identity. Hence, this is a third way consistent interacting theory of $(d-2)$-forms.
Also note that it has the standard $p$-form gauge symmetry $\delta B^I=d\xi^I$ where $\xi^I$ is a
$(d-3)$-form. 
We can generalize this equation  by including more terms in the shift \eqref{shift} that are
constructed out of $H^I$ and its derivatives. We may also add a matter current $J^I$ as
\be\label{eom2}
d\tilde H^I + \tfrac{1}{2}{\kappa}  \, \tilde H^J \wedge \tilde H^K\,f^I_{JK} = J^I\, ,  
\ee
provided that it satisfies
\be
d J^I = {\kappa}  \,  J^M \wedge \tilde H^N\,f^I_{MN} \, .
\ee
The current $J^I$ can be expressed in terms of a 1-form $y^I$ as
\be
J^I = -dy^I - \k f^I_{MN}  y^M \tilde H^N -\frac{1}{2} \k f^I_{MN} y^M \wedge y^N \, ,
\ee 
which can be seen by including $y^I$ in the shift \eqref{shift}.

\section{Future Directions}\label{future}

There are several interesting problems about this new class of theories some of which
we discuss below under three headlines.

\

{\it Constructing other examples:} All the known gravity examples \cite{Bergshoeff:2014pca, Ozkan:2018cxj, Afshar:2019npk}
are in three dimensions and the construction method of \cite{Afshar:2019npk} works only for {\it exotic} gravity models 
which have parity odd actions in the Chern-Simons like formulation but parity even metric field equations.
Adapting this method to parity violating models like the MMG \cite{Bergshoeff:2014pca} is desirable. Even more challenging is to find examples in higher dimensions. Recall that
the consistency of \eqref{mmg} works thanks to some very particular properties of the tensors $(S, C, J)$ which can schematically be listed as follows: i) $J\sim S^2$, ii) $\nabla J \sim CS$
since $C \sim \nabla S $, iii) $C$ is divergence free and part of the main field equation \eqref{mmg}, iv) $S. E_C=0$ where
$E_C$ is the $C$-field equation that is obtained from \eqref{mmg}. Finding analogues of these tensors in higher dimensions
seems difficult. Adapting the first order formulation of 3-dimensional higher derivative gravities  \cite{Hohm:2012vh, Bergshoeff:2014bia, Afshar:2014ffa, Merbis:2014vja} that we reviewed in section \ref{gravity} to other dimensions could be the key to achieve this. On the gauge theory side, it would be nice to find a YM covariant version of \eqref{eom1} where
unlike \eqref{gauged} gauge symmetries are compatible with each other. Using tensor hierarchies \cite{deWit:2008ta} might be useful for this purpose.
Finding an action with auxiliary fields for \eqref{eom1}
is another open problem.

\

{\it Supersymmetric extensions:} Until now only one example of a supersymmetric third way consistent model has been constructed \cite{Deger:2021fvv} which is the $N=1$ supersymmetric version of the 3-dimensional YM theory of \cite{Arvanitakis:2015oga} that we introduced in section \ref{gauge}. The starting point of \cite{Deger:2021fvv} is the $N=1$ off-shell supersymmetric topologically massive YM theory which contains a Majorana spinor field as the fermionic partner of the YM field. Its YM field equation looks exactly like \eqref{MYM} but instead of the
$F^2$ term on the r.h.s. there is a fermionic current $j_\m^I $. This current is conserved using the spinor field equation. By deforming the YM equation of this system with the $F^2$ term and demanding that under supersymmetry field equations are transformed to each other, the necessary extra terms in both equations are found. It is remarkable that this is possible without 
adding further bosonic terms to \eqref{MYM}. Moreover, it turns out that the end result can be obtained starting from the original system and performing a transformation of the form 
\be
A_\m^I \rightarrow A_\m^I + \a_1\tilde{F}_\m^I + \a_2j_\m^I \, ,
\ee
in the spinor field equation. The constants $\a_1$ and $\a_2$ are uniquely fixed by supersymmetry. The final YM equation can 
be obtained by a similar shift starting from $\tilde{F}_\m^I=0$.
Hence, the shift idea that already appeared in \eqref{shift_grav} and \eqref{shift} again plays a crucial role.
The construction of \cite{Deger:2021fvv} that we outlined above shows existence of a new class of supersymmetric theories that were not realized before and hence encourages many interesting directions to pursue such as trying to couple this model with other $N=1$ supersymmetric multiplets, constructing its extended supersymmetric versions, finding supersymmetric versions of gravity examples such as MMG \eqref{mmg} and $p$-form theories \eqref{eom1}.

\

{\it The third way and symmetry breaking:} Another way of understanding terms that lead to third way consistency is via symmetry breaking. In \cite{Chernyavsky:2020fqs} this was done
for the MMG by extending the Poincar\'e algebra to Hietarinta/Maxwell algebra \cite{Hietarinta:1975fu} which has one additional generator $Z_a$. Recall that the MMG auxiliary action \eqref{mmgaction} contains 3 fields $(e, \o, f_1)$ and they are now associated with the generators of this group, namely translations $P_a$, rotations $J_a$ and $Z_a$, respectively.  
Then, one can construct the Chern-Simons action
\be
S= \int <\O \wedge d\O + \frac{2}{3} \O\wedge \O \wedge \O> \, , 
\label{cs2}
\ee
for the Hietarinta/Maxwell algebra valued 1-form $\O=e^aP_a + \o^aJ_a + f_1^aZ_a$.  Spontaneous breaking of the
Hietarinta/Maxwell symmetry to its Poincar\'e sub-algebra gives rise to  extra terms in \eqref{cs2} and the
resulting action is equivalent to that of MMG \eqref{mmgaction} after some field redefinitions. It should be
possible to generalize this method for exotic gravity \cite{Ozkan:2018cxj, Afshar:2019npk}
and gauge theory examples \cite{Arvanitakis:2015oga, Broccoli:2021pvv}. The YM model of \cite{Arvanitakis:2015oga}
with $\m=0$ is already known to arise from a Higgs mechanism \cite{Mukhi:2011jp}. In \cite{Geiller:2019dpc} an alternative
description of MMG based only on dreibein and spin connection is given where the Lorentz gauge symmetry is broken. 
Finally, in \cite{Dereli:2019bom} it is shown that MMG is related with the break of the Weyl symmetry. 
It would be interesting to clarify connections between these different approaches.

\section*{Acknowledgements}
It is a great pleasure to dedicate this work to Prof. Tekin Dereli on the occasion of his 72nd birthday. I was fortunate to have
him as my master's supervisor, and I benefited from his support and guidance since then. With his distinguished scientific output
and leadership, he has been an inspiring example to many of us.

\

\noindent
I am grateful to ENS de Lyon, Laboratoire de Physique for hospitality and Research Fellowship Program of Embassy of France in Turkey for
financial support during the final phase of this study. I also would like to thank Henning Samtleben for helpful discussions.

\end{document}